%% file: cikm.tex
\documentclass[sigconf]{acmart}

\usepackage{tikz}
\usepackage{amsmath}

\usepackage{filecontents}
\usepackage{algorithm,algpseudocode}
\usepackage{graphicx}
\usepackage{amsmath}
\usepackage{subfigure}
\usepackage{multirow}

\usepackage{cleveref}
\usepackage{url}
 \usepackage[inline, shortlabels]{enumitem}
 \usepackage{makecell}
 \usepackage{listings}
 \usepackage{flushend}
 \usepackage{balance}
 \setlist[itemize]{align=parleft,left=2pt..1em}
\copyrightyear{2022} 
\acmYear{2022} 
\setcopyright{acmcopyright}\acmConference[CIKM '22]{Proceedings of the 31st ACM International Conference on Information and Knowledge Management}{October 17--21, 2022}{Atlanta, GA, USA}
\acmBooktitle{Proceedings of the 31st ACM International Conference on Information and Knowledge Management (CIKM '22), October 17--21, 2022, Atlanta, GA, USA}
\acmPrice{15.00}
\acmDOI{10.1145/3511808.3557139}
\acmISBN{978-1-4503-9236-5/22/10}
\begin{document}
\newcommand{\sysname}{SwiftPruner}
\newcommand{\lz}[1]{{\textcolor{red}{\it LZ: #1}}}
\newcommand{\yj}[1]{{\textcolor{blue}{\it YJ: #1}}}
\newcommand{\aw}[1]{{\textcolor{orange}{\it AW: #1}}}
\newcommand{\yh}[1]{{\textcolor{purple}{\it YH: #1}}}
\newcommand{\tc}[1]{{\textcolor{brown}{\it TC: #1}}}

\title{{\sysname}: Reinforced Evolutionary Pruning for  Efficient \\ Ad Relevance}


\author{Li Lyna Zhang}
\affiliation{%
  \institution{Microsoft Research}
  \country{lzhani@microsoft.com}
}

\author{Youkow Homma}
\affiliation{%
  \institution{Microsoft}
  \country{youkow.homma@microsoft.com}
}
\author{Yujing Wang}
\affiliation{%
  \institution{Microsoft}
  \country{yujwang@microsoft.com}
}

\author{Min Wu}
\affiliation{%
  \institution{Microsoft}
  \country{albw@microsoft.com}
}

\author{Mao Yang}
\affiliation{%
  \institution{Microsoft Research}
  \country{maoyang@microsoft.com}
}

\author{Ruofei Zhang}
\affiliation{%
  \institution{Microsoft}
  \country{bzhang@microsoft.com}
}
\author{Ting Cao}
\affiliation{%
  \institution{Microsoft Research}
  \country{ting.cao@microsoft.com}
}

\author{Wei Shen}
\authornote{Corresponding author}
\affiliation{%
  \institution{Microsoft}
  \country{sashen@microsoft.com}
}

\renewcommand{\shortauthors}{}
\renewcommand{\shorttitle}{ }
\input{abstract}

\ccsdesc[500]{Information systems~Information retrieval}
\ccsdesc[500]{Computing methodologies~Artificial intelligence}
\keywords{Transformer, Pruning, Neural Architecture Search, Ad Relevance}
\maketitle
\pagestyle{plain}
\input{intro}

\input{relatedwork}

\input{app}
\input{method}
\input{evaluation}

\input{conclusion}

{
	\bibliographystyle{ACM-Reference-Format}
	\bibliography{ref}
}
\end{document}

%% file: abstract.tex
\begin{abstract}

\noindent Ad relevance modeling plays a critical role in online advertising systems including Microsoft Bing. To leverage powerful transformers like BERT in this low-latency setting, many existing approaches perform ad-side computations offline. While efficient, these approaches are unable to serve cold start ads, resulting in poor relevance predictions for such ads. This work aims to design a new, low-latency BERT via structured pruning to empower real-time online inference for cold start ads relevance on a CPU platform. Our challenge is that previous methods typically prune all layers of the transformer to a high, uniform sparsity, thereby producing models which cannot achieve satisfactory inference speed with an acceptable accuracy. 


In this paper, we propose {\sysname} - an efficient framework that leverages evolution-based search to automatically find the best-performing \textit{layer-wise sparse} BERT model under the desired latency constraint. Different from existing evolution algorithms that conduct random mutations, we propose a reinforced mutator with a latency-aware multi-objective reward to conduct better mutations for efficiently searching the large  space of layer-wise sparse models. Extensive experiments demonstrate that our method consistently achieves higher ROC AUC and lower latency than the uniform sparse baseline and state-of-the-art search methods.
Remarkably, under our latency requirement of 1900us on CPU, {\sysname} achieves a \textbf{0.86\%} higher AUC than the state-of-the-art uniform sparse baseline for BERT-Mini on a large scale real-world dataset.
Online A/B testing shows that our model also achieves a significant 11.7\% cut in the ratio of defective cold start ads with satisfactory real-time serving latency.

\end{abstract}

%% file: intro.tex
\section{Introduction}

\noindent In sponsored search, ad relevance measures the semantic similarity between a user’s search query and an ad.  Given a user query, the system calculates its relevance with hundreds of potential ads to prevent irrelevant ads recommendation and to assist ranking and pricing of the most relevant ads. As such, ad relevance directly affects user and advertiser satisfaction with the product and serves as one of the most important tasks in online advertising. 

At Microsoft Bing, we previously released the TwinBERT~\cite{twinbert} and AutoADR~\cite{autoadr} models which leverage the latest advances in pre-trained language models (\textit{e.g.}, BERT)~\cite{transformer,bert} and neural architecture search (NAS) to effectively predict ad relevance. Due to the expensive inference cost of BERT, we decoupled the inferences for ad and query via a two-tower structure which consists of online query-side and offline ad-side sub-models. 
While this design avoids the time-infeasible online ad-side inference for hundreds of ads per query, the offline ad-side inference presents its own challenges. In particular, we encounter \textbf{coverage issues} due to \textit{cold start ads}, or \textit{cold ads} for short~\footnote{When the offline-computed ad-side data is not yet available during ad serving time online, we call such ads as cold ads. }: for a new ad, there is an inevitable delay between the time when it is available for online serving and when its offline-computed ad-side features are published online.


In this work, we aim to design a new semantic query-ad relevance model for cold ads which can be computed fully online on CPU under the real-time latency constraints. We therefore discard the two-tower structure and instead focus on \textbf{single-tower} BERT~\cite{bert} models where the query and ad content are concatenated as the model input.  Considering the real-time inference constraints, we select the state-of-the-art tiny transformer, BERT-Mini~\cite{bert-mini} as our model structure for cold ads, which achieves comparable accuracy with our previous models (\textit{i.e.}, TwinBERT and AutoADR) via knowledge distillation. Although optimized with both knowledge distillation~\cite{bert-mini,sanh2020distilbert} and quantization~\cite{onnxruntime_quantize,zafrir2019q8bert,kim2021bert}, the inference latency of BERT-Mini is still too high to be served online on our CPU-based platform. To afford the real-time inference for cold ads relevance, we call for new model compression methods to effectively reduce the latency of tiny transformers. 

Structured pruning methods have shown to be effective in reducing inference latency on standard hardware while mitigating accuracy loss. Previous works~\cite{nn_pruning,mccarley2019structured,kim2020fastformers,yao2021mlpruning,cofi} have demonstrated the large redundancy in transformers, but they focus on large transformers (e.g., BERT-Base and BERT-Large) and  simply set a uniform sparsity ratio for all layers.  However, we observe that individual attention and feed-forward (FFN) layers have different redundancy in tiny transformers (i.e. BERT-Mini). Pruning some critical encoder layers can lead to a significant accuracy drop while pruning other redundant layers can reduce latency without affecting the model accuracy. Moreover, attention and FFN layers contribute differently to the model latency. 
Therefore, we study layer-wise sparsity to achieve better trade-offs between accuracy and latency.


Layer-wise sparsity leads to a large design space to compromise between model accuracy and latency. While it is challenging to determine the sparsity for each layer from such a large space, NAS approaches such as reinforcement learning (RL)~\cite{pham2018efficient,zoph2017neural} and evolution search (EA)~\cite{real2019regularized,guo2020single}  have demonstrated effectiveness in automatically finding good architectures. Inspired by this, the questions naturally arise: \textit{Can we leverage NAS to search the space of layer-wise sparse models to find the best performing model under a given latency constraint? What kind of NAS algorithm is the most effective for our ad relevance scenario?}

To this end, we propose {\sysname}
, which automatically searches the best-performing layer-wise sparsity for structured pruning under a given latency constraint. Inspired by NAS, we formulate the pruning problem as a multi-objective optimization problem that considers both accuracy and inference latency of pruned transformers. First, we design a large search space that factorizes the BERT-Mini structure into layer-wise sparse configurations.
Then, an efficient search algorithm is employed to identify layer-wise sparse settings that can achieve both low latency and high accuracy. Finally, we select the pruned model with maximum validation Receiver Operating Characteristic (ROC) AUC under a certain latency constraint as the final model, named SwiftBERT.

Despite the success of EA and RL in NAS, both of them face specific challenges in our scenario. EA-based search shows simplicity and stability in many tasks~\cite{real2019regularized,guo2020single}. However, its evolution process is highly reliant on the mutation actions, and the search efficiency has no guarantee due to the uncontrollable random mutation. RL is effective but more time consuming than EA. Additionally, the stability of RL depends on careful hyper-parameter selection. Thus, it's difficult to tune for jointly optimizing accuracy and latency in our scenario. 

 To address the above limitations, we propose a reinforced evolution algorithm to search for the optimal layer-wise sparsity with a latency-aware multi-objective reward for transformers. Unlike traditional EA methods that randomly select a layer for sparsity mutation, our method introduces a reinforced mutator to learn \textit{(i)} which layer to mutate and \textit{(ii)} what sparsity value to assign to the mutated layer. Our search method leverages  the advantages of both EA and RL: we not only accelerate the search via reinforced mutations but also maintain the stability of EA.  
 Furthermore, to reduce the search cost caused by latency measurement, we build a latency predictor that can accurately predict the inference latency for sparse transformers.



We conduct extensive experiments with a real-world dataset for ad relevance under many latency constraints. For all latency constraints,  {\sysname} consistently achieves higher AUC and lower latency compared to the uniform sparsity baseline and state-of-the-art layer-wise search methods. Remarkably, {\sysname} can achieve a maximum of 43.46\% latency reduction on CPU for the BERT-Mini with a minimal AUC loss of 0.32\%. Compared to other search methods, {\sysname} demonstrates its superiority through higher AUC (up to +0.48\%) under all latency constraints, better search efficiency, as well as robustness to hyper-parameter selection. Finally, online A/B testing shows that SwiftBERT achieves a significant 11.7\% cut in the ratio of defective cold ads and 2.37\% increase in click-through-rate. This new model has been shipped into the Microsoft Bing ad relevance production model. 

We summarize the main contributions of our work as follows:
\begin{itemize}

	\item We propose {\sysname},
	an efficient framework that leverages NAS to search for the best-performing layer-wise sparse BERT under a desired latency constraint.   
\vspace{2px}
	\item We introduce a reinforced evolutionary search algorithm which efficiently predicts a layer and sparsity value for mutation. We demonstrate its superiority over EA and RL-based search methods in our scenario.
	\vspace{2px}
	\item We apply {\sysname} to ad relevance prediction and produce SwiftBERT models with various latency levels. Compared to uniform sparsity and other layer-wise search methods, SwiftBERT shows much higher AUC and lower inference latency. 
\vspace{2px}	
	\item After adding SwiftBERT to the ad relevance production pipeline, the ratio of defective cold ads is significantly reduced according to A/B testing results. To the best of our knowledge, SwiftBERT is the first single-tower, query-document crossing BERT architecture at Microsoft Bing that serves fully online within a real-time latency on CPU. Our method is also applicable to other scenarios such as dynamically generated ads~\cite{generate_search_text_ad, ad_gen} and search relevance~\cite{10.1145/2939672.2939677}.
\end{itemize}

%% file: relatedwork.tex
\section{Related Works}
\label{sec:relatedworks}


 

\noindent\textbf{Knowledge distillation}. 
Knowledge distillation \cite{sanh2020distilbert,sun2020mobilebert,jiao2020tinybert} is a powerful method which leverages scores, structures, or weights from a large, teacher BERT model to train a smaller, student BERT model.  While students that are 2-3$\times$ smaller than the teacher can be trained with minimal accuracy loss, Turc et al.~\cite{bert-mini} show that there is a steep drop-off in accuracy when distilling to tiny transformer students like those required in our scenario.  

\vspace{2px}
\noindent\textbf{Quantization}. Quantization ~\cite{zafrir2019q8bert,kim2021bert} of transformer models from FP32 to Int8 can produce models up to 4$\times$ as small with negligible accuracy loss regardless of the starting model size. While the model compression ratio from quantization is impressive, inference speedup often relies on specialized hardware~\cite{zafrir2019q8bert,chen2021quantization}.  On our  CPU deployment platform which is a standard Intel CPU, we find the benefit from model quantization is limited due to the lack of hardware support~\cite{onnxruntime_quantize}.

\vspace{2px}
\noindent\textbf{Pruning} can be categorized into: 1) unstructured and 2) structured pruning. Although unstructured methods~\cite{movement,gordon2020compressing,wang2020superglue} reach a relatively high sparsity ratio without significant accuracy drop, they yield very little actual latency benefits. Many common deployment platforms on modern hardware, particularly on CPU, often keep the original dense shapes of the uncompressed model
as it is difficult to leverage irregular sparse patterns for acceleration.

Structured pruning, on the other hand, allows for the removal of coherent groups of weights and can reduce latency without special hardware support. Early works ~\cite{michel2019sixteen,voita2019analyzing} showed that some attention heads in BERT can be removed without significant accuracy loss. Later, structured pruning was also expanded to act on both the attention and FFN layers of transformers~\cite{mccarley2019structured,kim2020fastformers,yao2021mlpruning,tprune}. Recently, Lagunas et al.~\cite{nn_pruning} introduced a hybrid pruning named nn\_pruning which prunes structured blocks in attention heads and paired rows/columns in FFN intermediate layers. When combined with movement pruning~\cite{movement}, which learns the importance score for weights during fine-tuning, nn\_pruning is able to achieve effective compression of BERT models.
However, these methods focus on maximum parameter removal by ranking the importance scores. They pre-define a target sparsity (compression ratio) and assign a uniform sparsity for all layers.  In our work, we find our production BERT models have different redundancy per layer and hence require layer-wise sparsity ratios.


\vspace{2px}
\noindent\textbf{Layer-wise sparsity}. In computer vision tasks, it has been demonstrated that different CNN layers have different redundancy, and rule- or heuristic-based unstructured pruning~\cite{lee2021layeradaptive,sun2021dominosearch,evci2021rigging} may be used to find the appropriate layer-wise sparsities to reduce their redundancy. AMC~\cite{amc} is the first work that leveraged reinforcement learning (RL) to automatically search the optimal layer-wise sparsity for CNNs. To accelerate exploration, AMC conducts weight magnitude pruning. However, this method is less effective for fine-tuned BERT models on downstream tasks~\cite{movement}. Unlike these previous works, we focus on pruning BERT models for NLP tasks.  In our work, we propose reinforced evolutionary search, which is more efficient than RL. During the search, we conduct movement-based structured pruning rather than magnitude pruning. Therefore, we can efficiently find the optimal layer-wise sparsity for a target latency constraint with minimal accuracy loss.

\vspace{2px}
\noindent\textbf{AutoML}. Hyper-parameter optimization (HPO) and neural architecture search (NAS) are both prevalent approaches for conducting efficient optimization within a large search space. The Tree Parzen Estimator (TPE)~\cite{tpe} algorithm is one of the most popular methods for hyper-parameter optimization which utilizes Bayesian optimization to search high-dimensional conditional spaces. However, TPE relies on the quality of initial samples and has limitations in robustness~\cite{NIPS2012_05311655}. On the other hand, reinforcement learning (RL) and evolution search (EA) have achieved state-of-the-art performance in NAS tasks. Works focusing on RL-based search~\cite{pham2018efficient,zoph2017neural} train an RNN controller with reinforcement learning to determine a sequence of operators and connections that specify the architecture of a neural network. One limitation of RL-based search is the need for careful hyper-parameter selection to guarantee stability. In contrast, EA is less sensitive to hyper-parameter selection. Real et al. ~\cite{real2019regularized} introduces aging evolution, which samples a best parent candidate at each iteration and then conducts random mutations to create children candidates. While EA addresses RL's shortcomings, EA's search efficiency cannot be guaranteed due to its heavy reliance on random mutation. 

In our work, we propose a reinforced RNN controller to conduct predictable mutation which greatly accelerates EA's search efficiency. The idea of reinforced mutations on EA has been previously investigated in RENAS~\cite{chen2019renas}, but they focus on architecture search for CV tasks without considering inference latency. To the best of our knowledge, our method is the first to propose and apply this idea to prune NLP models to a desired latency. 


%% file: app.tex
\section{Ad Relevance Scenario}
\label{sec:app}

\subsection{Background}

\noindent In sponsored search systems, the user submits a \textit{query} to the search engine, and the engine delivers \textit{ads} along with the search results. The ads shown to the user are chosen based on a number of parameters including \textit{relevance} between the query and ad, pricing, and user behavior. When ads are more relevant to the user's query, the user is more satisfied with the search product and the advertiser receives increased value from their bid.  Thus, ad relevance serves as a cornerstone to a healthy sponsored search ecosystem.

The Microsoft Bing ad relevance system consists of a cascade of models which feed into one another to make many ad relevance-based decisions in the ad stack.  These decisions include filtering out the most irrelevant ads, placing relevant ads higher in the search results, and pricing ads based on their relevance.  One of the primary goals of the ad relevance system is to minimize the number of \textit{defective} ad impressions.  An ad is determined to be defective for a query if their semantic similarity is low as determined by a set of pre-determined guidelines.  As such, some of the most important models in the ad relevance system are the query-ad \textbf{semantic relevance models}.  These semantic models take the query and ad text as input and produce a prediction of defectiveness, called the \textbf{semantic score}, which is then used by other models.


\begin{figure}[t]
	\centering
	\includegraphics[width=0.95\columnwidth]{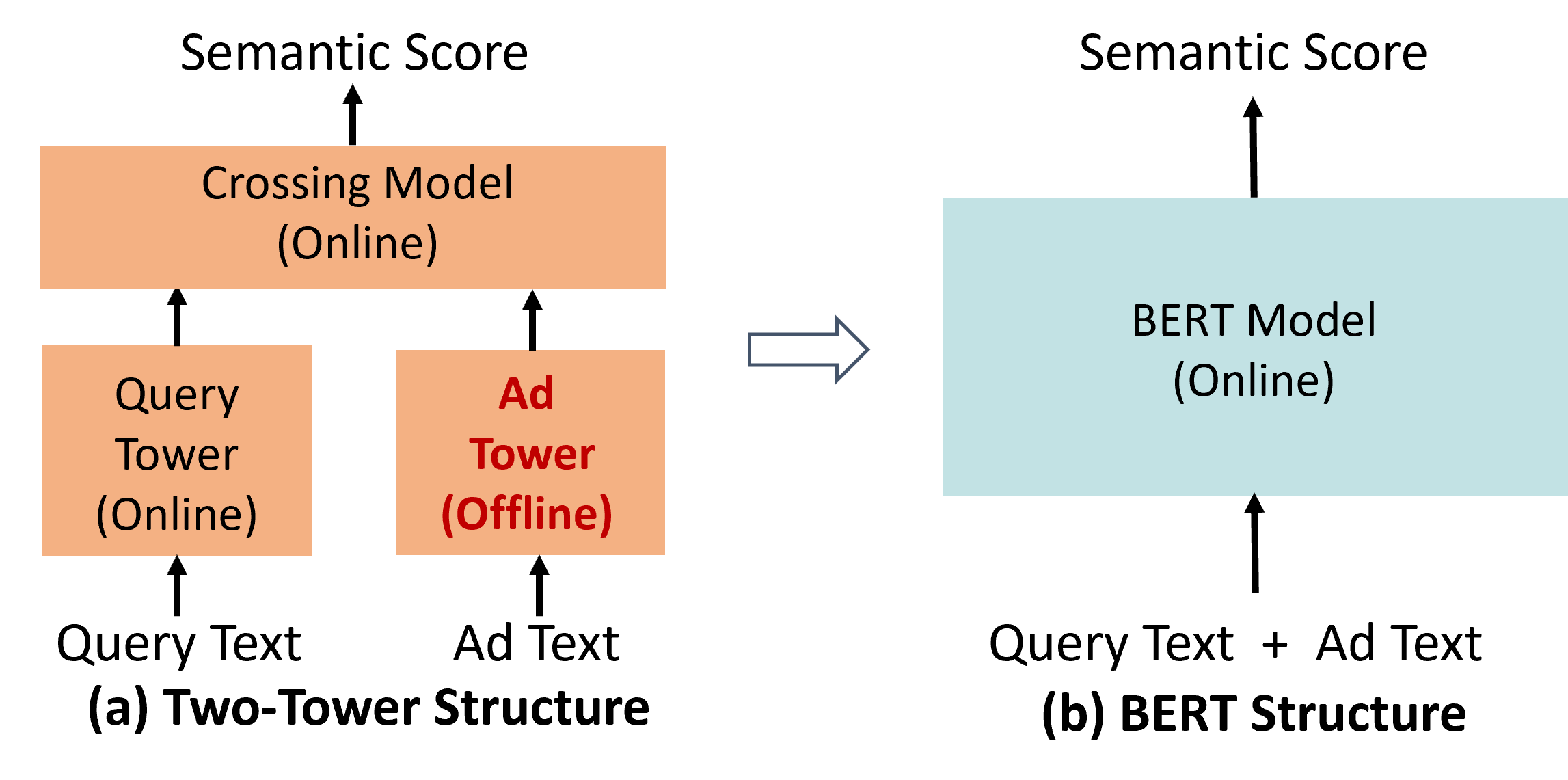}	
	\caption{Comparison of the hybrid offline-online two-tower structures of TwinBERT and AutoADR (a) vs the fully online BERT structure of SwiftBERT (b). }
	\label{fig:modelstructures}
\end{figure}

Currently, our system contains a few semantic relevance models with the most powerful being TwinBERT~\cite{twinbert} and AutoADR~\cite{autoadr} which leverage the latest advances in transformers.  These two models share a similar two-tower structure as shown in Fig.~\ref{fig:modelstructures}(a) which decouples the query and ad inferences for efficient computation. The query-side model serves online and is computed just once per user query. On the other hand, the ad-side model is inferenced offline for the billions of servable ads, and the embedding outputs are then published to the online service.  Given a candidate query-ad pair, the semantic score is then computed online by running a small crossing model on the query and ad tower outputs.

The two-tower structure allows us to use large, powerful models in our production system; however, it also introduces a challenge.  When advertisers create new, servable ads, there is a delay between when the ad can be shown to users and when the ad-side tower computation is published to the online system.  This delay results in ads that are missing their offline-computed ad-side data at the online serving stage.  We call such ads, \textbf{cold ads}.  Conversely, we call ads that have their ad-side data online, hot ads.  For cold ads, our ad relevance system currently relies on weak semantic relevance models that do not leverage transformers.  As a result, millions of defective cold ads pollute the system daily and degrade the overall quality of ads shown to users.  To address these defective cold ads, we call for a new, powerful transformer-based semantic relevance model which can be used to specifically serve cold ads.

\subsection{Serving Cold Ad Relevance}




\noindent In order to compute a semantic score for cold ads, we cannot use offline components in the model. We therefore discard the two-tower structure and instead opt for a fully online, single-tower BERT where the input is a concatenation of the query and ad text as shown in Fig.~\ref{fig:modelstructures}(b).  While this single-tower structure addresses the coverage issue, a new challenge arises. For each query-hot ad pair, the two-tower structure allows us to run just a small crossing model online for each query-ad pair to get their semantic score. This crossing model can be run on hundreds of ads per query with minimal impact on system latency.  On the other hand, for every query-cold ad pair, we must now run an entire BERT model online. Inferencing even a very small BERT model on tens of ads per query has a measurable impact on overall system latency.  Therefore, our new cold ads model must be designed to run within a strict latency budget.  Specifically, we find that the new single tower model must run in less than 1900us per query-cold ad pair on our platform.

\begin{table}[t]
	\begin{center}
		\small
		\begin{tabular}	{c|c|c|c}
			\hline
			Model &Size&AUC (\%)& Latency (us) \\
			\hline
			BERT Teacher Model & 2B & 88.98 & - \\
			\hline 
			BERT-Mini (4L-256H)~\cite{bert-mini} &11.3M &87.15 &3274.24  \\
			2L-256H~\cite{bert-mini} &9.7M &85.16 & 1727.67 \\
			4L-128H~\cite{bert-mini} &4.8M &84.23 & 1841.29  \\
			\hline
		\end{tabular}
		\caption{AUC and latency of distilled models from~\cite{bert-mini}.  }
		\label{tbl:distill}
	\end{center}
\end{table}

To start, we distill some of the smallest, state-of-the-art tiny transformers  from Turc et al.~\cite{bert-mini} which meet our system latency budget of 1900us.  The chosen models are a 2-layer, 256-hidden dimensional BERT (2L-256H) and a 4-layer, 128-hidden dimensional BERT (4L-128H).  After distilling the two models using our powerful teacher model, we find there is a large gap in ROC AUC between these two students and the teacher as shown in Table~\ref{tbl:distill}.

To strike a better balance between prediction performance and latency, we examine the next largest model checkpoint from ~\cite{bert-mini}: BERT-Mini, which is a 4-layer, 256-hidden dimensional BERT model with 4 attention heads and 1024 FFN dimensions.  While BERT-Mini achieves significantly better performance as shown in Table~\ref{tbl:distill}, its inference time is over 70\% above the latency limit.  In order to retain its good performance while decreasing the latency, we explore different model compression methods.  We first apply Int8 quantization~\cite{zafrir2019q8bert,kim2021bert} on our CPU deployment platform, but the inference latency is still too high beyond our requirement.
We next turn to pruning to close the remaining latency gap.  In the following sections, we study how our {\sysname} method can be applied to BERT-Mini to generate a new, low-latency semantic relevance model for cold ads, which we call SwiftBERT.

%% file: method.tex
\section{Methodology}

\begin{figure*}[t]
	\centering
	\includegraphics[width=1\textwidth]{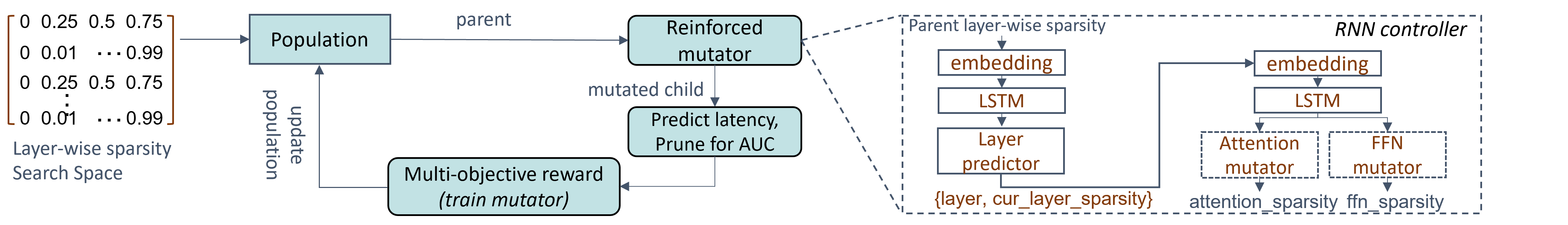}	
	\caption{ Overview of {\sysname} for searching the optimal layer-wise sparse transformers.}
	\label{fig:arch}
\end{figure*}

\noindent In this section, we present the
 {\sysname}  for layer-wise pruning of tiny transformers (i.e., BERT-Mini) such that the pruned model meets the given latency constraints.

We formulate our problem as:
\begin{equation}
	\label{eq:problem}
	\begin{aligned}
		\displaystyle (a_1, f_1, ... a_l, f_l)^*=\mathop{\arg\min}\limits_{a_1, f_1,... a_l, f_l}{L}(A(a_1, f_1, ... a_l, f_l))
		\\
		\mathrm{ s.t. }\;   LAT (A(a_1, f_1, ... a_l, f_l)^*)<=T,
	\end{aligned} 
\end{equation}
where $A(\cdot)$ is the transformer model with a specific sparsity setting, and $a_i$ and $f_i$ are the sparsity ratios for the attention and FFN subnetworks in the $i^{th}$ encoder layer, respectively. {\sysname} aims to find the optimal layer-wise sparsity
$(a_1, f_1, ... a_l, f_l)^*$ from the $1^{st}$  to  $l^{th}$ encoder layer that has the minimum cross-entropy loss, $L$, (corresponding to the maximum accuracy) while the latency (denoted by $LAT$) also meets the constraint, $T$. 

To achieve this, we first design a fine-grained sparsity search space which contains a large amount of models that can meet various latency constraints after pruning. Then we introduce a reinforced evolutionary search
algorithm equipped with a latency predictor. Instead of random mutation, the algorithm learns to mutate a layer with predicted sparsity towards the best accuracy-latency trade-off. Fig.~\ref{fig:arch} illustrates the overall pipeline. 

\begin{table}[t]
	\begin{center}
		\small
		
		\begin{tabular}	{c|c|c|c}
			\hline
	     Layer & Base config.& Sparsity space&Prune pattern \\
	     \hline 
	     MSA ($a$) &4 & (0, 0.25, 0.5, 0.75)& head \\
	     FFN ($f$) & 1024&(0, 0.01, 0.02, ...0.99) &dimension\\
	     \hline
		\end{tabular}
		\caption{Candidate sparsity for each encoder layer. The total search space contains $\sim$2$\times$10$^{10}$ settings for pruning.}
		\label{tbl:search_space}
	\end{center}
\end{table}
\subsection{Search Space Design}

\noindent Unlike previous NAS works~\cite{chen2019renas,nasbert} that search the optimal operator for each layer, our goal is to search for the optimal sparsity setting for each layer of the transformer model. Since our deployment platform has no support for accelerating unstructured pruned models, we design a large transformer search space for structured pruning, as detailed in Table ~\ref{tbl:search_space}. 
Following previous works~\cite{michel2019sixteen,mccarley2019structured, nn_pruning,cofi} which show that the majority of attention heads and FFN intermediate dimensions can be pruned with minimal accuracy loss, we also prune the attention layers via head pruning and FFN layers via dimension pruning. This hybrid pruning granularity setting has demonstrated a better accuracy-efficiency trade-off~\cite{cofi,nn_pruning}. Concretely, we use the entire head as the smallest pruning granularity in attention. The FFN layer contains two paired linear layers, so we prune the rows in the first layer (i.e., the intermediate layer output) and the corresponding columns in the second linear layer (i.e., the second layer input).

\vspace{3px}
\noindent\textbf{Head sparsity}.  For an attention layer with $H$ heads, we allow pruning of \{0,1..,H-1\} heads (at least one head is retained to avoid pruning the entire layer).  Since our base model, BERT-Mini~\cite{bert-mini}, has 4 attention heads in one encoder layer, we have 4 candidate settings of sparsity ratio: 0, 0.25, 0.5 and 0.75, which correspond to the pruning of 0, 1, 2 and 3 heads, respectively. 
 
\vspace{3px}
\noindent\textbf{FFN sparsity}. The original FFN intermediate dimension size $F$ of BERT-Mini is 1024. A full search space would contain all possible dimension sizes, $F$, but would also lead to search space explosion for larger dimensions. Fortunately, we observe that there are negligible latency differences between $F$ and $F+1$ on our platform. Therefore, we reduce the search space of dimensions to 100 candidates, where different sparsity leads to noticeable latency change. As shown in Table~\ref{tbl:search_space}, we allow FFN layer sparsity ratios to range from 0 to 0.99 with a step size of 0.01 (i.e., a step of $\sim$10 in dimension size ). 

In total, for our 4 encoder-layer SwiftBERT model, the search space contains $\sim$2$\times$10$^{10}$ candidate settings, which is extremely large 
  and poses practical challenges for traditional NAS search algorithms. If a candidate model could be evaluated in a few seconds via weight sharing and performance prediction~\cite{ofa,donna,attentivenas}, a naive evolution algorithm could quickly find good performing models by evaluating thousands of candidates. However, evaluating the AUC for layer-wise settings requires considerable training cost on our large-scale real-world dataset. Unlike in NAS tasks, introducing weight sharing in pruning poses difficulty in accurate AUC evaluation.
 Therefore, we require a higher efficiency search algorithm that can find the optimal layer-wise sparse setting in a few hundred search trials.
  In the following sub-section, we introduce our proposed search method to tackle this problem.


\subsection{Reinforced Aging Evolution}
\label{sec:reinforced_evolution}

\algrenewcommand{\algorithmiccomment}[1]{$\triangleright$ #1}
\begin{algorithm}[h]
	\small
	\caption{Reinforced Aging Evolution Algorithm}\label{alg}
	\begin{flushleft}
		\textbf{Input}: population size $P$, total number of models to explore $N$, sample \\size $S$, latency\_auc tradeoff $\alpha$, target latency $T$\\
		
	\end{flushleft}
	\begin{algorithmic}[1]

		\State $population^{(0)}$ $\gets$ \texttt{initialize} ($P$)
		\State $history$ $\gets$ $population^{(0)}$	
		
		\For{$i$=1: ($N$-$P$)}
		\State $parent$ $\gets$ \texttt{sample\_with\_maximum\_reward} ($population^{(i-1)}$,$S$)
		\State $child$ $\gets$ \texttt{reinforced-mutate} ($parent$)
		\State $reward$ $\gets$ \texttt{get\_reward} ($child$, $T$,$\alpha$ ) (see Equation~\ref{eq:reward})
		\State update mutator with maximizing $reward$ by REINFORCE~\cite{reinforce} 
		\State $population^{(i)}$$\gets$ add $child$ to right of $population^{(i-1)}$, remove the oldest model from left of  $population^{(i-1)}$
		\State add $child$ to $history$
		\EndFor
		\State return maximum-AUC model under the $T$ from the $history$
	\end{algorithmic}
\end{algorithm}

\noindent To efficiently search the large space of candidates, we propose the reinforced aging evolution as shown in Fig.~\ref{fig:arch}. 
Algorithm~\ref{alg} formulates the overall procedure of {\sysname}. Following the original aging evolution, we first randomly initialize a population of $P$ models from the search space, where each model is encoded with a specific layer-wise sparsity setting. After this, evolution improves the initial population in mutation iterations (line 3). At each iteration, we sample $S$ random models from the population and select the model with maximum reward as the $parent$. 
We then run the reinforced mutator to mutate a layer sparsity on the $parent$ and construct a new model, called the $child$. Once the $child$ is constructed, we compute its reward by Equation~\ref{eq:reward}, which considers both accuracy and inference latency. We train the reinforced mutator by maximizing the reward with REINFORCE~\cite{reinforce}. At the end of each iteration, we remove the oldest model in the current population and add the $child$ to the population for the next iteration. After the evolution process finishes, we collect all the explored models that are under the target latency constraint $T$ and select the model with maximum validation AUC as the final model. 



\vspace{3px}
\noindent\textbf{Reinforced mutator}. We now introduce the key component in reinforced aging evolution, namely the reinforced mutator, which conducts sparsity mutations on the sampled $parent$ model. In the original aging evolution algorithm, the mutator randomly samples one layer and performs a random modification of the sparsity ratio for this layer. However, as described in Section~\ref{sec:relatedworks}, uncontrollable mutation cannot guarantee the efficiency of optimizing our reward. 

Instead of random mutations, the reinforced mutator is designed to 1) predict a layer to mutate, and 2) predict a sparsity ratio for the selected layer, so that the mutated $child$ has a better chance to achieve the optimal AUC-latency trade-off. Fig.~\ref{fig:arch} (right) illustrates the architecture of our reinforced mutator. Guided by the two prediction targets, we design an RNN controller which sequentially selects a new layer to mutate and the corresponding new sparsity ratio. The controller is implemented as a recurrent neural network consisting of two sub-networks. 

In the first sub-network, we take the layer-wise sparsity of the $parent$ as input. The sub-network starts with an embedding layer with a bidirectional LSTM to learn the effect of each layer-wise sparsity ratio of the $parent$. Based on the output of the LSTM, we apply a fully connected layer with \textit{Softmax} as the layer predictor. The predictor estimates the mutation probability of each layer, and we sample a layer $l_i$ for mutation according to the probability. 

After determining the layer $l_i$ for mutation, we run the second sub-network to predict a new sparsity ratio for this layer. We take the predicted layer index $l_i$ and its current sparsity ratio $f_i$ of the $parent$ as the input to this model. As shown in Fig.~\ref{fig:arch} (right), another LSTM layer is leveraged to learn the effect of changing the sparsity ratio. Since attention and FFN layers have different sparsity ratios (shown in Table~\ref{tbl:search_space}), we design two different fully-connected classification layers for attention and FFN layers, called the \textit{Attention mutator} and \textit{FFN mutator}, respectively.
The corresponding classification branch is selected according to the type of layer $l_i$. For instance, if layer $l_i$ is an attention layer, we forward the input to the \textit{Attention mutator} branch to predict a new sparsity ratio, $f_i^*$.

\vspace{3px}
\noindent\textbf{Latency-aware reward}. The objective of our reinforced mutator is to maximize the latency-aware reward as follows:

\begin{equation}
\label{eq:reward}
\begin{aligned}
\displaystyle reward= AUC(m) \times   \left[ \frac{LAT(m)}{T} \right ]^w
\end{aligned}
\end{equation}
where $AUC(m)$ denotes the AUC of model $m$, $LAT(m)$ denotes the inference latency on the target platform, and $T$ is the target latency.  $w$ is the weight factor defined as:
\begin{equation}
\label{eq:tradeoff}
\begin{aligned}
  w=\begin{cases}
 0, & \text{if}\; LAT(m)\leq T\\
 \alpha, & \text{otherwise}
 \end{cases}
\end{aligned}
\end{equation}
In our scenario, we have a hard limit on the pruned model latency. 
 Therefore, we set $w=0$ if the latency of the searched model $m$ is less than $T$ to find the layer-wise sparsity with the highest AUC under constraint, $T$. Otherwise, we set $w=\alpha$ to penalize the objective value and discourage mutations that violate the latency constraint. $\alpha$ is a negative number, and in our experiment, we empirically set $\alpha=-1$. 

 The computation of AUC and latency in the reward function is time-consuming. To eliminate the cost of latency measurement, we build a latency predictor that can accurately predict the latency for layer-wise sparse models. To reduce AUC computation cost, we conduct fast structured pruning on a small training set. We detail these optimizations below.  

\vspace{3px}
\noindent\textbf{Latency prediction}. Measuring the latency for a model  requires deploying the model on the target platform and then running the model many times on the same input. As such, it would be time-consuming to measure the candidate model latency at each iteration of the reinforced aging evolution algorithm. 
In lieu of this, we build a latency predictor which can accurately predict the latency for each layer-wise sparse model on our deployment platform. 
First, we randomly sample layer-wise sparse models from our search space and measure the latency on the target platform. We then train a random forest regression model~\cite{randomforest} to predict the latency values from the sampled models' layer-wise sparsity ratios.

\vspace{3px}

\noindent\textbf{Structured pruning}. 
We conduct structured pruning to obtain the AUC in the reward function. It's non-trivial to select a structured pruning algorithm for our scenario. In our application, the base model BERT-Mini is first initialized with pretraining weights~\cite{bert-mini} and then fine-tuned on our real-world training set. Since movement pruning~\cite{movement} has demonstrated the advances in removing unimportant neurons for fine-tuning pretrained models, we leverage Hugging Face's nn\_pruning library~\cite{nn_pruning}, which is an improved version of movement pruning to support structured pruning. 

The original nn\_pruning conducts block pruning in the attention layer.  To determine which blocks are pruned, it divides weight matrices $W_q$ (Query $Q$), $W_k$ (Key $K$), $W_v$ (Value $V$), and $W_o$ (output weight) into individual blocks. For instance, when the block size equals the head size, 
the Key  matrix $W_k$ in the 4-head BERT-Mini will be divided into $W_{k1}$, $W_{k2}$, $W_{k3}$, $W_{k4}$.  
 Each block is assigned with a score, which is used to measure the importance for  block weights to final model accuracy. 
 However, since the importance scores for each head's $W_{ki}$, $W_{qi}$, $W_{vi}$ are  learned independently, there is no guarantee to prune an entire head. Therefore, we cannot directly apply block pruning in our scenario. For example, when the sparsity is set to prune 1 head,  block pruning  removes one block in each $W_q$, $W_k$, $W_v$.  Based on the importance scores, we might prune the first head by $W_{q1}$, the second head by $W_{k2}$ and the third head by $W_{v3}$. In this case, no head would get removed and there would be no latency reduction.

 To solve this problem, we apply a simple but effective method. For each head $i$, we  constrain its corresponding four matrices $W_{qi}$, $W_{ki}$, $W_{vi}$ and $W_{oi}$ with one shared importance score, so that all matrices in one head will be pruned at the same time. 
 We evaluate the two different implementations on pruning the same model (i.e., the output model with 80\% sparsity using the original nn\_pruning in Fig.~\ref{fig:layerwise_uniform}), and they achieve comparable AUC scores.  

Additionally, 
to further reduce the search cost, we randomly sample a small subset of the training data which is sufficiently large enough to evaluate model performance. As a result, we can quickly get the approximate AUC for ranking a candidate model  in 4 minutes on a 4-V100 GPU node.

%% file: evaluation.tex
\section{Evaluation}
\label{sec:evaluation}
\subsection{Dataset and Setting}

\noindent\textbf{Dataset}. The training set for our study is a large-scale sample of real impressed query-ad pairs from Microsoft Bing's search log. Since it would be laborious to manually label each query-ad pair in the training set as defective/non-defective, we follow the practice in AutoADR~\cite{autoadr} of generating the defective probability for each sample using a powerful BERT teacher model.  The teacher model itself is trained on millions of human-classified query-ad pairs. All of our pruned models are trained on the binary cross entropy loss between the teacher score and model score.  The test set in our experiments contains 100k query-ad pairs which are classified as defective/non-defective by human judges, and we report the ROC AUC of the model scores on this set.  The teacher has an AUC of 88.98\% on the test set.


\vspace{2px}
\noindent\textbf{Latency measurement}. To build the latency predictor, we randomly sample 5000 models from our sparsity search space and measure their inference latency on a Intel(R) Core(TM) i7-7700 CPU device.  The inference framework is ONNX runtime 1.8.2. To make full utilization of ONNX optimizations,  we apply quantization~\cite{onnxruntime_quantize} from FP32 to Int8 as well as ONNX transformer optimizations~\cite{onnxruntime_operatorfuse} for all models. 
We report the average latency of 1000 runs with an input sequence length of 38 and batch size of 1. The sampled layer-wise sparsity and measured latency pairs are  split 8:2 between the training and validation sets for the latency prediction model. The RMSE is only 70us and RMSPE is 3.88\% on the validation data. This suggests that the latency prediction model can be used to replace the expensive latency measurement with little error introduced.

\vspace{3px}
\noindent\textbf{Search algorithm details.} In our experiment, we explore $N$=500 models in total. To set the population size $P$ and sample size $S$, 
we consider 20, 50 and 100 as in  ~\cite{real2019regularized,chen2019renas}. The  results show there is no obvious difference after training to converge. We finally set the  $P$ and $S$ to 50. The initial populations are randomly sampled from our search space with a relaxed latency constraint (i.e., 1.15$\times$ of the target constraint). For the mutation controller, the embedding size is 104 (4 candidates in Attention and 100 candidates in FFN), and the hidden state size of the 2-layer LSTM is 100. We use the Adam optimizer with a learning rate of 0.001.  

At each mutation step, the controller samples a model, and we calculate the latency-aware multi-objective reward for this mutation action. As introduced in Section~\ref{sec:reinforced_evolution}, we predict the latency for the sampled model using our latency predictor and conduct layer-wise pruning to get the ROC AUC score. To reduce the search cost incurred during pruning, we prune on a subset of 500 mini-batches, which is randomly sampled from our full training set. The AUC is evaluated on a small validation set for the search algorithm.  When the mutation process finishes, we collect all models that fulfill the latency constraint, and select the model with maximum validation AUC. In our experiment, the search process can finish within 24 hours on a 4-V100 GPU node. 

\vspace{3px}
\noindent\textbf{Final layer-wise pruning and evaluation.} For the final model setting selected by our search algorithm, we conduct layer-wise pruning using the full large-scale training set. Concretely, we run 3 epochs of pruning and 1 epoch of fine-tuning with a batch size of 8192 and initial learning rate of 3e-5.  The other settings and hyper-parameters are kept the same as nn\_pruning\footnote{The code of nn\_pruning is at: \url{https://github.com/huggingface/nn\_pruning}}~\cite{nn_pruning}. We report the final ROC AUC on our test set.

\subsection{{\sysname} Under Various Latency Constraints}

\begin{figure}[t]
	\centering
	\includegraphics[width=1\columnwidth]{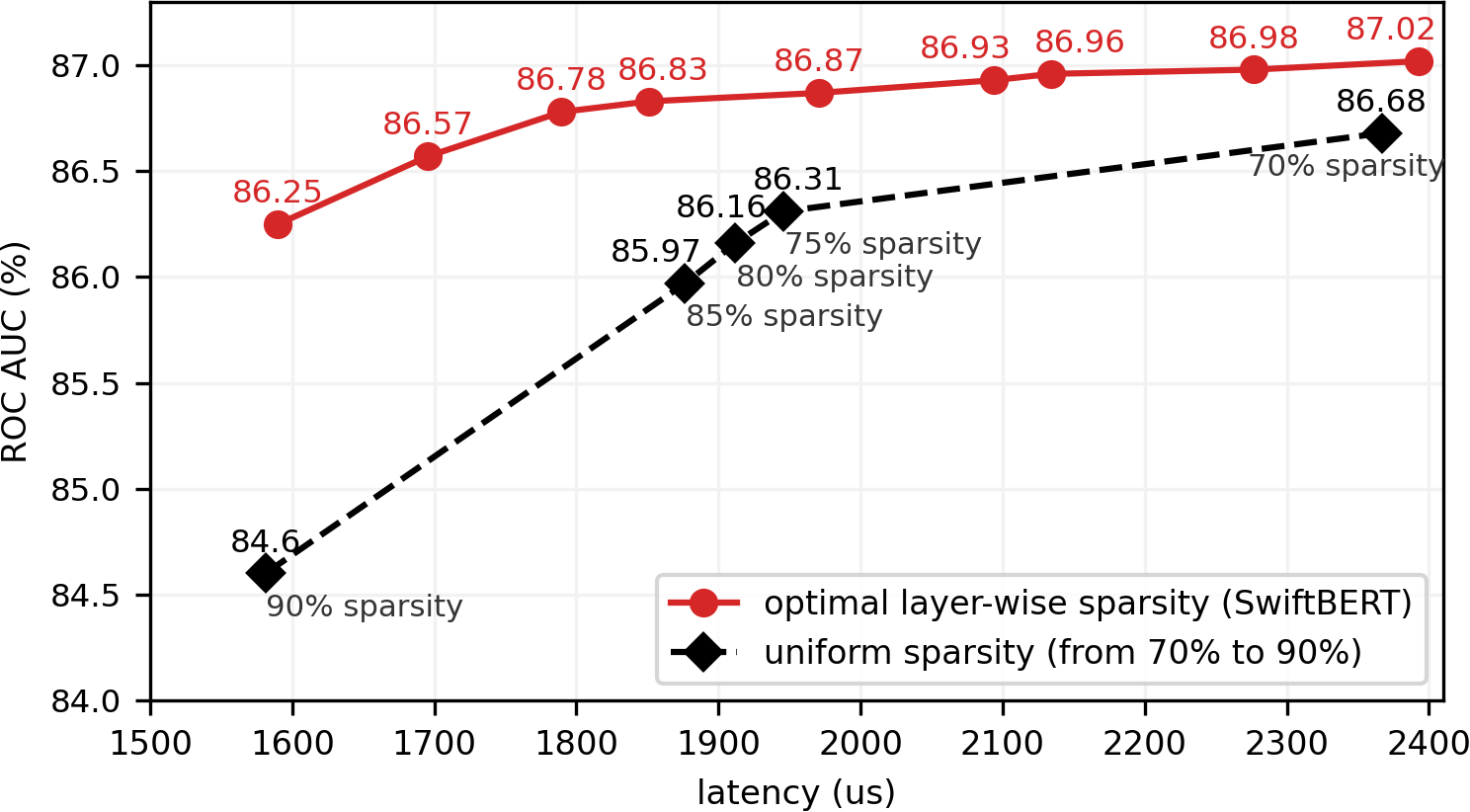}	
	\caption{Compared to  uniform sparsity pruning, {\sysname} consistently outputs the best model settings under various latency constraints.  Moreover, {\sysname} is able to construct models for a wide range of latency constraints whereas uniform sparsity model latencies are fragmented and difficult to tune to a specific latency requirement. }
	\label{fig:layerwise_uniform}
\end{figure}

\begin{table}[t]
	\begin{center}
		\small
		
		\begin{tabular}	{c|c|c|c}
			\hline
		Model & Size& AUC (\%)&Latency \\
			\hline 
		 	BERT-Mini~\cite{bert-mini} &11.3M & 87.15 & 3274.24 us\\
		 	\hline
		 	uniform pruning~\cite{nn_pruning} (85\% sparsity) & 8.86M & 85.97& 1876.44 us\\
		 	SwiftBERT & 8.99M& \textbf{86.83}& \textbf{1851.16 us}\\
			\hline
		\end{tabular}
		\caption{End-to-end performance under the required 1900us latency constraint. SwiftBERT achieves +0.86\% better AUC than uniform pruned model with smaller latency.}
		\label{tbl:end-to-end}
	\end{center}
\end{table}
\noindent\textbf{Setup}. We now report our reinforced evolutionary pruning performance and compare 
with the state-of-the-art uniform sparsity method, nn\_pruning~\cite{nn_pruning} by Hugging Face.  Since nn\_pruning does not support latency-aware pruning, we set multiple sparsity ratios for comparison. Following the original paper, we apply hybrid-filled pruning for the best AUC and efficiency trade-off and use square block (32$\times$32) pruning on the attention layer and dimension pruning on the FFN layer.  A fine-tuning is conducted after the pruning to re-fill the 0-valued neurons in the pruned model. For fair comparison, we run 3 epochs of pruning and 1 epoch of fine-tuning under the same random seed.      To demonstrate the effectiveness of {\sysname}, we set a wide range of  latency constraints from 1600us to 2400us.

\vspace{2px}
\noindent\textbf{Results and Analysis}.  Fig.~\ref{fig:layerwise_uniform} gives a full comparison of models with different compression methods and latency constraints. 
Our results show that by searching the optimal layer-wise sparsity, {\sysname} yields both significant latency and AUC improvement compared to the uniform sparsity baseline. 
 More specifically, we observe the following: (i) layer-wise sparsity demonstrates the superiority in removing redundancy in tiny transformers compared to uniform sparsity. Under the same-level latency, {\sysname} consistently outperforms uniform sparsity pruning. Specifically, SwiftBERT (1851.16us) achieves +0.86\% higher AUC than the 85\% uniform sparsity model (1876.44us). The advantages of layer-wise sparsity becomes more competitive under extremely low latency constraints.  SwiftBERT (1590.00us) outperforms the 90\% uniform sparsity model (1581.05us) by a significant +1.65\% higher AUC.
 (ii) Our method demonstrates the effectiveness in pruning tiny transformer to various latency constraints, which is crucial in practical deployment. In Fig.~\ref{fig:layerwise_uniform}, we can see that the final latency is not linearly reduced by the increase in uniform sparsity. In particular, it's hard to tune a sparsity to obtain a model with around 1700us latency via nn\_pruning. Meanwhile, {\sysname} can prune a model while retaining the maximum AUC under all latency constraints.
 
 In our online system, the latency of SwiftBERT is constrained to 1900us. Table~\ref{tbl:end-to-end} suggests that we can even achieve much lower latency than the 1900us constraint with a satisfactory ROC AUC. Remarkably, our searched layer-wise sparsity model (1851.16us) can significantly reduce the original BERT-Mini's latency by 43.46\% with an acceptable AUC loss of just 0.32\%.  The uniform pruned model, on the other hand, has a 1.18\% AUC drop under the 1900us constraint, which demonstrates the efficiency of layer-wise sparsity in this extremely low-latency regime.

\subsection{Comparison of Search Methods}

\noindent The previous section demonstrated that our searched layer-wise sparsity configuration consistently outperforms uniform sparsity. In this section, we compare {\sysname} with other methods that are also applicable to search layer-wise sparse models.

\vspace{2px}
\noindent\textbf{Comparison baselines}. We implement three other baselines, which are representative of the state-of-the-art in model architecture and hyper-parameter search. 
\begin{itemize}

	\item Tree Parzen Estimator (TPE). We apply the TPE algorithm~\cite{tpe} by formulating the problem as a hyper-parameter search where the hyper-parameters are the sparsity values for the attention heads and FFN  of the 4 encoder layers. 
	\item Reinforcement Learning (RL). Our implementation follows the well-known ENAS~\cite{pham2018efficient}. We design and train an RNN controller to predict the sparsity for each layer. The  controller consists of 4 LSTM sub-networks, where each sub-network predicts the attention and FFN sparsity for a corresponding encoder layer. 

	\item Aging Evolution (EA). Our implementation follows the original  algorithm in ~\cite{real2019regularized}. EA is conducted with the same setting to {\sysname}, except that the mutation action is made by randomly sampling a layer and its new sparsity setting.
\end{itemize}

For fair comparison,  we use one training recipe and hyper-parameter setting for all methods. Specifically, all three baselines search for the same number of 500 models. For EA, the population size $P$ and sample size $S$ are the same with ours (i.e., set to 50). All search methods optimize the same reward function in Equation~\ref{eq:reward}.  

\begin{figure}[t]
	\centering
	\includegraphics[width=0.98\columnwidth]{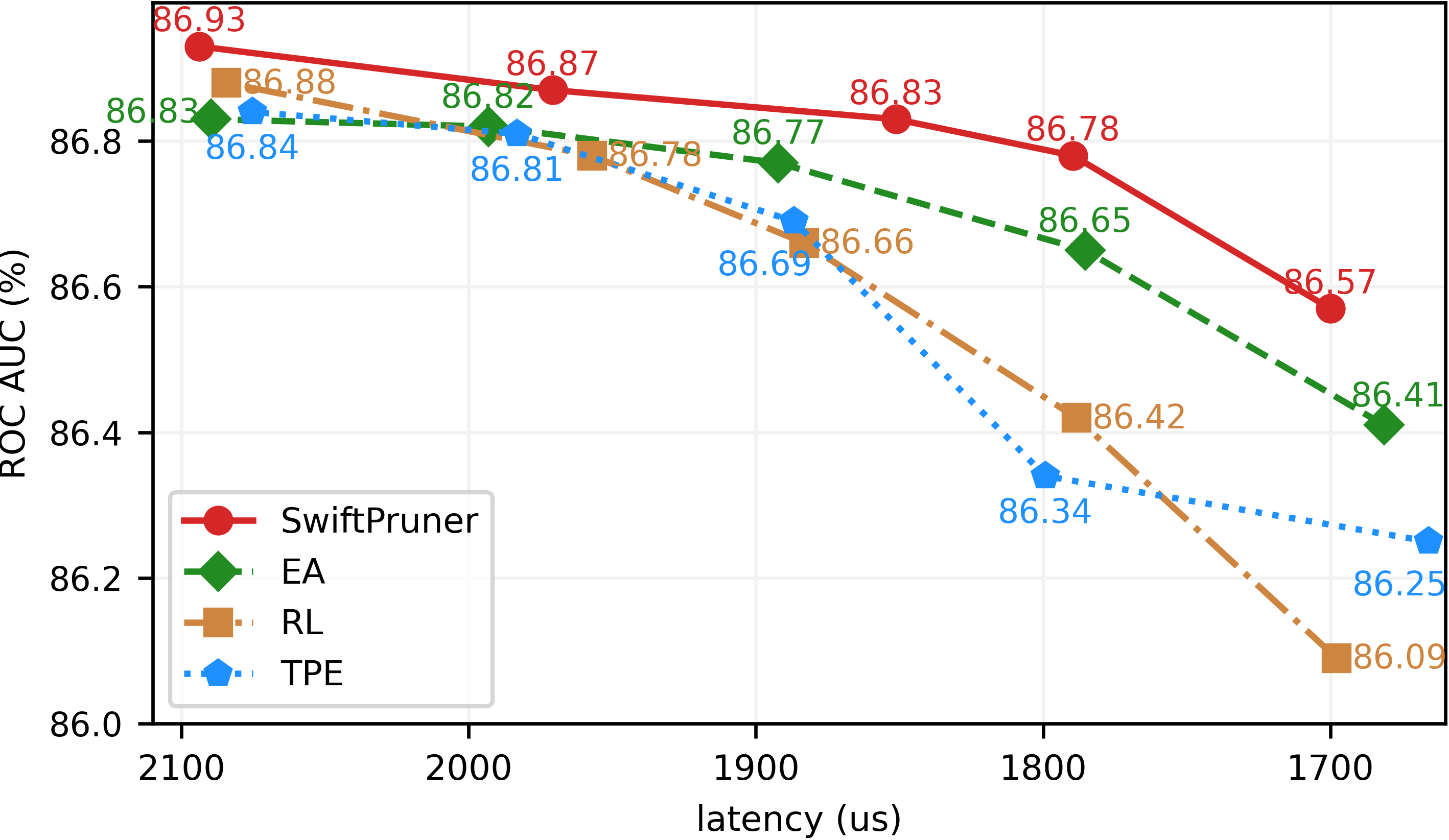}	
	\caption{Comparison of search methods under different latency constraints, $\alpha=-1$. The x-axis is decreasing in latency.}
	\label{fig:search_method}
\end{figure}
\vspace{2px}
\noindent\textbf{Search results comparison}. We first compare the final AUC for the searched models.
 Fig.~\ref{fig:search_method} reports comparisons between {\sysname} and the other search methods  under five latency constraints (1700us - 2100us). Compared to the three baselines, {\sysname} consistently reaches the highest AUC under all latency constraints. Specifically, we improve the AUC by 0.1\% (EA), 0.09\% (TPE), and 0.05\% (RL)  under the maximum 2100us constraint. We also observe that the AUC improvements by {\sysname} are even higher under lower latency constraints. Significantly, under the smallest 1700us constraint, we outperform EA, TPE, and RL by 0.16\%, 0.32\%, and 0.48\% higher AUC, respectively.

\begin{figure}[t]
	\centering
	\includegraphics[width=0.9\columnwidth]{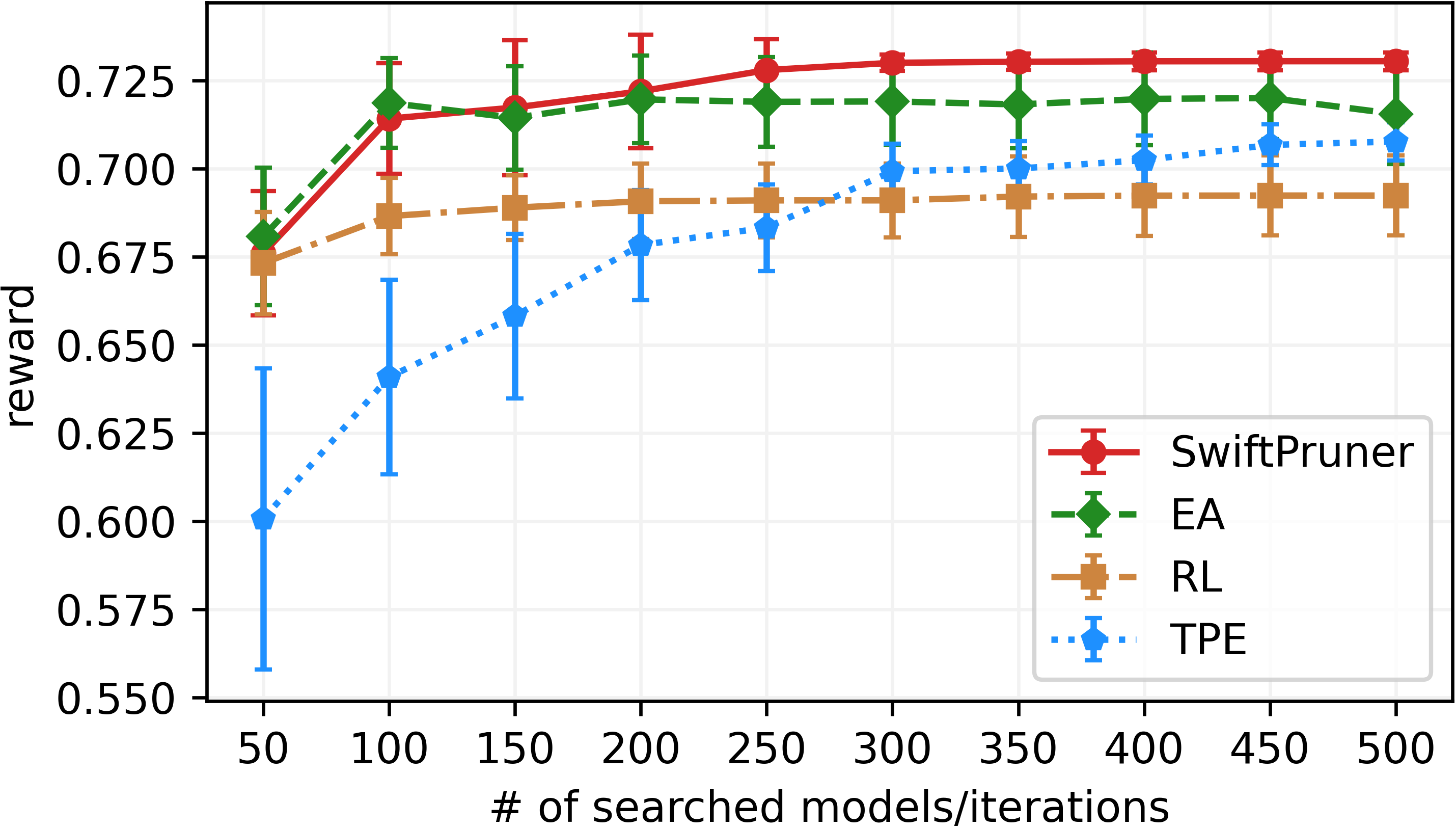}	
	\caption{Search efficiency comparison under the same search iteration ($\alpha=-1$).}
	\label{fig:reward_comparsion}
\end{figure}

\vspace{2px}
\noindent\textbf{Search efficiency comparison}. For a deeper understanding of various search methods, we not only compare the AUC scores of the best searched models, but also compare the search efficiency. Since all the search methods optimize both latency and AUC, we compare the achieved reward (computed by Equation~\ref{eq:reward}) with trade-off ratio $\alpha=-1$.
Fig.~\ref{fig:reward_comparsion} compares the search efficiency of {\sysname} to EA, RL and TPE under the 1900us constraint. For each method, we keep track of the searched models for every 50 iterations. Since EA and {\sysname} are evolution-based methods, we evaluate them by the reward mean and variance of models in the population (i.e., 50 models). 
For TPE and RL, we select the top 50 models with the best reward over time. 

As shown in Fig.~\ref{fig:reward_comparsion}, {\sysname} achieves better efficiency than EA, RL and TPE. Compared to TPE and RL, {\sysname} consistently achieves higher rewards all the time. Compared to EA, {\sysname} degrades a little over the first few iterations of reinforced mutation, but it quickly surpasses EA after 150 iterations.

\vspace{2px}
\noindent\textbf{The impact of AUC-latency trade-off $\alpha$}. $\alpha$ controls the AUC-latency trade-off during the search process. A smaller $\alpha$ indicates a larger penalty on the latency. To study the impact of different $\alpha$ on search methods, we set three $\alpha$: -0.3, -0.7 and -1. We run {\sysname} and our baselines with different $\alpha$ under the same 1900us constraint.

Table~\ref{tbl:latency_ratio_alpha} summarizes the searched AUC  under different $\alpha$. From the results, we observe that: (i) for each $\alpha$, {\sysname} reaches the highest AUC; (ii) while TPE and RL AUC vary significantly across the three $\alpha$ settings, {\sysname} and EA have stable performance with little variance. This suggests that evolution-based algorithms are more robust to hyper-parameter selection.

\begin{table}[t]
	\begin{center}
		
		\begin{tabular}	{c|c|c|c|c}
			\hline
			$\alpha$ & TPE& RL&EA & SwiftPruner \\
			\hline 
		  -0.3 &86.71 &86.65 & 86.75& \textbf{86.82}\\
		  -0.7 &86.79 &86.30 &86.76 & \textbf{86.84}\\
		  -1 &86.69 & 86.66&86.77 &\textbf{86.83}\\
			\hline
		\end{tabular}
		\caption{AUC (\%) achieved by different latency-AUC tradeoff $\alpha$. We set the same 1900us constraint for all methods.}
		\label{tbl:latency_ratio_alpha}
	\end{center}
\end{table}


In summary, we demonstrate that {\sysname} outperforms other search methods with higher AUC models, better search efficiency and greater robustness to hyper-parameter selection.

\subsection{Pruned Model Visualization}

\begin{figure}[t]
	\centering
	\includegraphics[width=0.99\columnwidth]{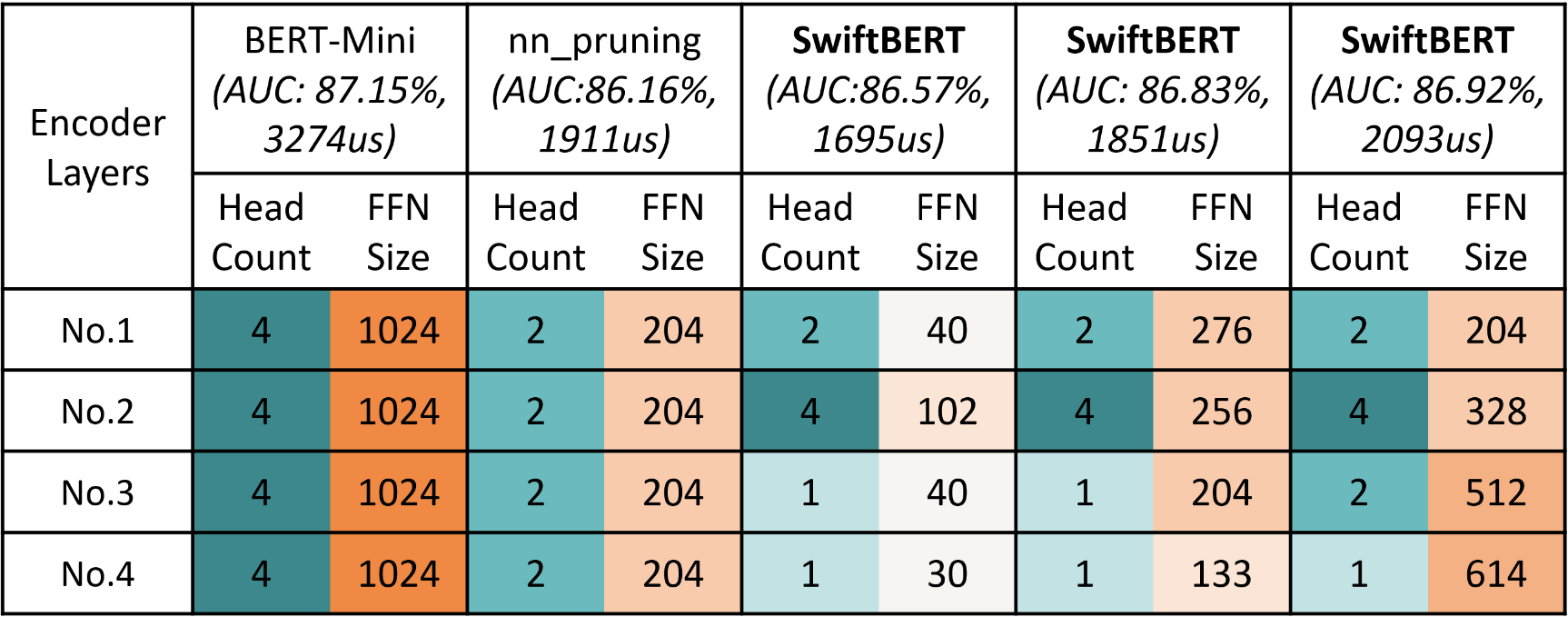}	
	\caption{Visualization of pruned model structure. Insights for ad relevance: first two encoders prefer larger attention head count; FFN  has more redundancy than attention. }
	\label{fig:visualize}
\end{figure}

\noindent We now visualize the SwiftBERT structures and identify some heuristics for designing more efficient pruned BERT architectures.  By doing so, we hope to motivate and inspire further study of layer-wise pruning and architecture design of transformer models.


Fig.~\ref{fig:visualize} shows the pruned structures of both nn\_pruning and {\sysname}. We observe that deeper layers have more redundancy than lower layers in BERT-Mini, which is consistent with the observations in CNN models~\cite{metapruning}. Compared to uniform sparsity, our layer-wise sparsity method prunes less in crucial layers and prunes more in less-crucial layers. For instance, while nn\_pruning prunes 2 heads in each encoder layer, our method automatically learns to keep all attention heads in the 2nd encoder layer and prunes 3 heads in the 3rd and 4th layers. As a result,  SwiftBERT (1851us) has a comparable latency with nn\_pruning but a much higher AUC (+0.67\%). Moreover, we observe that {\sysname} learns to keep more attention heads and prune more FFN intermediate layers under tighter latency constraints. For example, SwiftBERT (1695us) retains the same attention structure as SwiftBERT (1851us), but keeps only 0.03\%-0.1\% of the original FFN intermediate dimensions. 


\subsection{Impact on Ad Production}

\noindent In this section, we study the impact of our offline AUC gain by integrating the SwiftBERT model with latency constraint 1900us (SwiftBERT 1900) into the Microsoft Bing ad relevance system.  To do so, we add SwiftBERT 1900 as a new query-ad semantic relevance model for cold ads whose semantic score is then used to inform decisions later in the ad stack as described in Section~\ref{sec:app}. 

\begin{table}[t]
\begin{center}
\begin{tabular}{c|c|c}
\hline
Metric    & Cold ads   & Cold+Hot ads  \\ \hline
Defect Rate &  -11.7\%  &  -3.3\%   \\
RPM        &  +7.94\%   & +0.27\% \\
CTR             & +2.37\% & +0.65\%  \\
QBR                & -4.43\%   & -0.16\% \\
Avg. Latency       & - & +0.10\%  \\
99.9th \% Latency  & -  & +1.50\%  \\ \hline

\end{tabular}
\caption{Online KPI impact of adding SwiftBERT 1900 to serve as a new cold ads semantic relevance model. The online test shows significant positive KPI's in both  cold ads segment and on all (cold+hot) ads with minimal latency impact.}
\label{tbl:onlineab}
\end{center}
\end{table}

For the task of filtering defective cold ads, we find SwiftBERT 1900's semantic score can lift the relevance system's defective cold ad classifier performance by a significant +0.96\% in ROC AUC.  

To measure the other effects of adding SwiftBERT 1900 to the relevance system, we conduct an online A/B test and study the impact on both the cold ads and overall (i.e., hot+cold ads) segments. The KPI results from our online A/B test is outlined in Table~\ref{tbl:onlineab} and a description of each metric is provided below.

\noindent\textbf{Defect Rate:} Ratio of defective ads which are shown to users.  One major goal of the ad relevance system is to minimize this rate.

\noindent\textbf{Revenue per Mille (RPM):} This metric is the revenue per 1000 queries.  Increasing this value while maintaining or improving quality of ads represents an improvement in the overall ad system.

\noindent\textbf{Click-through-rate (CTR):} This is the average number of clicks per ad impression.  CTR tends to be correlated with the defect rate and positive trends signal an improvement in the quality of ads shown to users.

\noindent\textbf{Quick back rate (QBR):} This is the percent of clicks that result in users ``quickly'' returning to the search page, indicating they are unsatisfied with the clicked ads.  QBR tends to correlate with the defect rate and reduced QBR suggests ad quality improvements.

\noindent\textbf{Avg. Latency:} Average latency of the system to process all ads per query.  We would like our models to not impact the system latency significantly. 

\noindent\textbf{99.9th\% Latency:} Latency of the system to process all ads for the slowest 0.1-percentile of queries.  We would like to keep this number small such that even the slowest queries with the largest ad loads do not time out.

In Table~\ref{tbl:onlineab}, we see a strict improvement in every KPI in both the cold and overall segments.  In particular, we observe a great 11.7\% cut in cold ad Defect Rate which translates to a 3.3\% cut in overall ad Defect Rate.  Furthermore, we see strong positive metrics in user behavior with a 2.37\% increase in CTR and a substantial 4.43\% cut in QBR for cold ads delivered.  Finally, the online A/B test shows a greater revenue after adding SwiftBERT 1900 with a 7.94\% RPM increase among cold ads and a 0.27\% RPM increase among hot+cold ads.  Since we add a new semantic model which runs at real time for all cold ads at online serving stage, the latency of the system increases, as expected. However, due to our efficient pruning methodology, the latency impact of running SwiftBERT 1900 for all cold ads is reasonably small with just a 1.50\% increase in 99.9th\% latency. This means even after adding SwiftBERT 1900, our system can continue to serve user queries with high ad load without timing out.  

In summary, our SwiftBERT 1900 model delivers significant performance improvements in ads relevance, user behavior metrics, and revenue while adding minimal latency to the system.

%% file: conclusion.tex
\vspace{-4px}
\section{Conclusion and Future works}
\noindent In this paper, we introduced {\sysname}, a novel end-to-end framework that searches the best-performing layer-wise sparse BERT models  under a desired latency constraint. {\sysname} provides an efficient reinforced evolutionary search algorithm, which learns to conduct effective mutations rather than random ones. Extensive experiments demonstrate its effectiveness and efficiency compared to other search methods. Real-world online testing shows that our pruned model, 
SwiftBERT achieves a significant defective cold ads reduction and enables real-time, transformer-based online serving for cold ads on CPU at Microsoft Bing.  



Finally, {\sysname}'s fast online inference provides benefits beyond just the cold ads scenario. We have also successfully deployed {\sysname} to serve relevance in dynamic ads~\cite{generate_search_text_ad, ad_gen} where offline ad-side computation is infeasible because ad text is dynamically generated online.  Additionally, we plan to apply {\sysname} to search relevance~\cite{10.1145/2939672.2939677} and vertical ads in the future.

\balance